\documentclass[11pt]{article}%
\usepackage{amsfonts}
\usepackage{amsmath}
\usepackage{amssymb}
\usepackage{graphicx}%
\providecommand{\U}[1]{\protect\rule{.1in}{.1in}}
\newtheorem{theorem}{Theorem}
\newtheorem{acknowledgement}[theorem]{Acknowledgement}

\begin{document}

\title{On the Dependence of Quantum States on the Value of Planck's Constant}
\author{Maurice de Gosson\thanks{maurice.de.gosson@univie.ac.at}\\University of Vienna\\Faculty of Mathematics (NuHAG)\\Oskar-Morgenstern-Platz 1, 1090 Vienna AUSTRIA
\and Makan Mohageg\thanks{makan.mohageg@csun.edu}\\CSUN-College of Physics and Astronomy\\18111 Nordhoff Street, Northridge, CA 91330}
\maketitle

\begin{abstract}
We begin by discussing known theoretical results about the sensitivity of
quantum states to changes in the value of Planck's constant $h$. These
questions are related to positivity issues for self-adjoint trace class
operators, which are not yet fully understood. We thereafter briefly discuss
the implementation of experimental procedure to detect possible fluctuations
of $h$.

\end{abstract}

\section{Introduction}

There has been for a long time an ongoing and controversial debate on whether
the fine structure constant $\alpha$ is really constant; its non-constancy
would imply that at least one of the related quantities $e,c,h$ could also be
variable; see for instance \cite{Barrow1,BarrowWebb,Duff}. So far all attempts
to test the variability of physical constants have relied on experimental
evidence. In the present paper we propose a theoretical approach for detecting
possible variations of Planck's constant; in principle this method would be
able to detect such a variation no matter how small it is. In the present
Letter we examine some of the consequences of the possible non-constancy of
$h$ would have for pure mixed and quantum states; the mathematical theory is
far from complete, and related to difficult questions involving positivity
issues, so we will limit ourselves mainly to the case of Gaussian pure or
mixed states which is well understood. We thereafter suggest experimental
procedures in the form of \textit{Gedankenexperimente}.

\section{The Density Matrix and the KLM\ Conditions}

As soon as one deepens the elementary description of mixed states one is
confronted with difficult (and yet not totally solved) mathematical problems
which are usually ignored in experimental physics. This difficulty is due to
the relation between density matrices and their associated Wigner
distributions, and which is not as straightforward as it could seem at first
sight. It lies in the verification of \textit{positivity issues}, and these
very much depend on the value which is given to Planck's constant. Recall that
a mixed quantum state in $\mathbb{R}^{n}$ is characterized by its density
matrix
\begin{equation}
\widehat{\rho}=\sum_{j}\lambda_{j}|\psi_{j}\rangle\langle\psi_{j}|
\label{density}%
\end{equation}
where $\lambda_{j}\geq0$, $\sum_{j}\lambda_{j}=1$, and $\langle\psi_{j}%
|\psi_{j}\rangle=1$. The operator $|\psi_{j}\rangle\langle\psi_{j}|$ is the
orthogonal projection in $L^{2}(\mathbb{R}^{n})$ on the state $|\psi
_{j}\rangle$. The datum of $\widehat{\rho}$ is equivalent to that of Wigner
distribution function (WDF)%
\begin{equation}
W_{\widehat{\rho}}(x,p)=\left(  \tfrac{1}{2\pi\hbar}\right)  ^{n}\int
e^{-ipy/\hbar}\langle x+\tfrac{1}{2}y|\widehat{\rho}|x-\tfrac{1}{2}y\rangle
d^{n}y \label{wigdensitybraket}%
\end{equation}
(see \textit{e.g}. Hillery \textit{et al}. \cite{hillery}, Littlejohn
\cite{Littlejohn}) that is, using formula (\ref{density})%
\begin{equation}
W_{\widehat{\rho}}(x,p)=\rho(x,p)=\sum_{j}\lambda_{j}W\psi_{j}(x,p)
\label{wigdensity}%
\end{equation}
where $W\psi_{j}$ is the usual Wigner transform of $\psi_{j}$:
\begin{equation}
W\psi_{j}(x,p)=\left(  \tfrac{1}{2\pi\hbar}\right)  ^{n}\int e^{-ipy/\hbar
}\psi_{j}(x+\tfrac{1}{2}y)\psi_{j}^{\ast}(x-\tfrac{1}{2}y)d^{n}y.
\label{wigtrans}%
\end{equation}
We address the following question:

\begin{quotation}
\textit{Suppose we are given a real function} $\rho$ \textit{on}
$\mathbb{R}^{2n}$\textit{. How can we know whether this function is the Wigner
distribution of some density operator} $\widehat{\rho}$, \textit{and what
happens if we replace} $\hbar$ \textit{with another real number}
$\hbar^{\prime}>0$\textit{, that is, if we allow Planck's constant to vary?}
\end{quotation}

The key to an understanding of this problem lies in the following remark: in
mathematical terms a density operator (or matrix) on $L^{2}(\mathbb{R}^{n})$
is a self-adjoint trace class operator $\widehat{\rho}$ with unit trace, and
which is in addition positive semidefinite: $\widehat{\rho}\geq0$ (we will say
for short \textquotedblleft positive\textquotedblright). The last condition
means that we have
\begin{equation}
\langle\psi|\widehat{\rho}|\psi\rangle\geq0\text{ \ \textit{for all} }\psi\in
L^{2}(\mathbb{R}^{n}) \label{positivity}%
\end{equation}
(this condition implies the self-adjointness of $\widehat{\rho}$ since
$L^{2}(\mathbb{R}^{n})$ \ is an infinite-dimensional Hilbert space, but we
keep it since we are going to deal with self-adjoint trace-class operators
$\widehat{\rho}$ which are not necessarily positive). Such operators
$\widehat{\rho}$ are compact, and the spectral theorem then implies that we
can always write $\widehat{\rho}$ in the form (\ref{density}). The definition
(\ref{wigdensity}) of the Wigner distribution then shows that if we view
$\widehat{\rho}$ as a Weyl operator (which always is possible
\cite{Birk,Birkbis}) then $\rho=W_{\widehat{\rho}}$ is just the Weyl symbol of
$\widehat{\rho}$ divided by $(2\pi\hslash)^{n}$ (\textit{loc.cit.}).

\section{Gaussian States}

Consider first the simple example of a centered normal probability
distribution on $\mathbb{R}^{2}$ defined by
\begin{equation}
\rho_{X,P}(x,p)=\frac{1}{2\pi\sigma_{X}\sigma_{P}}\exp\left[  -\frac{1}%
{2}\left(  \frac{x^{2}}{\sigma_{X}^{2}}+\frac{p^{2}}{\sigma_{P}^{2}}\right)
\right]  \label{Gauss1}%
\end{equation}
where $\sigma_{X},\sigma_{P}>0$; the variances are $\sigma_{X}^{2}$ and
$\sigma_{P}^{2}$ and the covariance is $\sigma_{XP}=0$. It is known
\cite{Narcow3,Narconnell} that $\rho_{X,P}$ is the Wigner distribution of a
density operator $\widehat{\rho}_{X,P}$ if and only if it satisfies the
Heisenberg inequality $\sigma_{X}\sigma_{P}\geq\frac{1}{2}\hbar$. If we have
$\sigma_{X}\sigma_{P}=\frac{1}{2}\hbar$ (which we assume from now on) then
$\rho_{X,P}$ is the Wigner distribution of the coherent state
\[
\psi_{X}(x)=(2\pi\sigma^{2})^{-1/4}e^{-x^{2}/2\sigma_{X}^{2}}%
\]
and $\widehat{\rho}_{X,P}$ is then just the pure-state density operator
$|\psi_{X}\rangle\langle\psi_{X}|$, whose Wigner distribution is precisely
$\rho_{X,P}=W\psi_{X}$. Suppose now that we replace $\hbar$ with a number
$\hbar^{\prime}>0$ playing the role of a \textquotedblleft
new\textquotedblright\ Planck's constant. If we still want $\rho_{X,P}$ to
qualify as the Wigner distribution of a quantum state the spreading
$\sigma_{X}$ and $\sigma_{P}$ must satisfy the new Heisenberg inequality
$\sigma_{X}\sigma_{P}\geq\frac{1}{2}\hbar^{\prime}$; since the product
$\sigma_{X}\sigma_{P}$ is fixed from the beginning as being $\frac{1}{2}\hbar$
this implies that we must have $\hbar^{\prime}\leq\hbar$ . This means that if
we decrease the value of Planck's constant, then $\widehat{\rho}_{X,P}$
becomes the density operator of a now mixed quantum state, but if we increase
its value, then the Gaussian $\rho_{X,P}$ cannot be a Wigner distribution, but
is a probability density representing a classical state. (Intuitively, the
decrease of Planck's constant has the effect of making the Gaussian $\rho$ too
sharply peaked around the origin, which causes the violation of the Heisenberg inequality.)

The discussion above extends without difficulty to generalized Gaussians
\[
\rho_{\Sigma}(z)=(2\pi)^{-n}\sqrt{\det\Sigma^{-1}}e^{-\frac{1}{2}\Sigma
^{-1}z^{2}}%
\]
in $2n$-dimensional phase space; $\Sigma$ is a positive-definite symmetric
$2n\times2n$ matrix which is identified with the covariance matrix:
\begin{equation}
\Sigma=\int(z-\bar{z})(z-\bar{z})^{T}\rho(z)d^{n}z\text{ \ , \ }\bar{z}=\int
z\rho(z)d^{n}z \label{covmat}%
\end{equation}
(we use the notation $z=(x,p)$, $x$ and $p$ being the generalized coordinate
vectors $(x_{1},...x_{n})$ and $(p_{1},...p_{n})$ viewed as column vectors in
all calculations). A necessary and sufficient condition for $\rho_{\Sigma}$ to
be the Wigner distribution of a density matrix is the following (see
\cite{Narcow3,Narcow,Narconnell}):
\begin{equation}
\Sigma+\frac{i\hbar}{2}J\text{ \textit{is positive semidefinite}}
\label{cond1}%
\end{equation}
(for short: $\Sigma+\frac{i\hbar}{2}J\geq0$) where $J=%
\begin{pmatrix}
0 & I\\
-I & 0
\end{pmatrix}
$ is the standard symplectic matrix. It can be restated as a condition on the
eigenvalues of $J\Sigma$: noting that these are the same as the eigenvalues of
the antisymmetric matrix $\Sigma^{1/2}J\Sigma^{1/2}$ they must be of the type
$\pm i\lambda_{1},...,\pm i\lambda_{n}$ with $\lambda_{j}>0$; the numbers
$\lambda_{1},...,\lambda_{n}$ are the \textit{symplectic eigenvalues} of
$\Sigma$. Condition (\ref{cond1}) is then equivalent to $\lambda_{j}\geq
\tfrac{1}{2}\hbar$ \ for $j=1,2,...,n$, that is to%
\begin{equation}
\lambda_{\min}\geq\tfrac{1}{2}\hbar\label{cond3}%
\end{equation}
where $\lambda_{\min}$ is the smallest symplectic eigenvalue. Notice that
(\ref{cond3}) reduces to the Heisenberg inequality $\sigma_{X}\sigma_{P}%
\geq\frac{1}{2}\hbar$ when
\[
\Sigma=%
\begin{pmatrix}
\sigma_{x_{j}}^{2} & 0\\
0 & \sigma_{p_{j}}^{2}%
\end{pmatrix}
\]
as in the example above. More generally, writing $\Sigma$ in block-form $%
\begin{pmatrix}
\Sigma_{xx} & \Sigma_{xp}\\
\Sigma_{px} & \Sigma_{pp}%
\end{pmatrix}
$ where $\Sigma_{xx}=(\sigma_{x_{j},x_{k}})_{1\leq j,k\leq n}$, $\Sigma
_{xp}=(\sigma_{x_{j},p_{k}})_{1\leq j,k\leq n}$ and so on, one can show
\cite{Birk,Birkbis} that the conditions (\ref{cond1}), (\ref{cond3}) are
equivalent to the Robertson--Schr\"{o}dinger inequalities (RSI)
\begin{equation}
\sigma_{x_{j}}^{2}\sigma_{p_{j}}^{2}\geq\sigma_{x_{j},p_{j}}^{2}+\tfrac{1}%
{4}\hbar^{2}. \label{RS}%
\end{equation}
Assume that the inequalities in (\ref{cond3}) all become equalities:
$\lambda_{1}=\cdot\cdot\cdot=\lambda_{n}=\tfrac{1}{2}\hbar$. The RSI are then
saturated, that is
\begin{equation}
\sigma_{x_{j}}^{2}\sigma_{p_{j}}^{2}=\sigma_{x_{j},p_{j}}^{2}+\tfrac{1}%
{4}\hbar^{2} \label{RSbis}%
\end{equation}
and the state is now a generalized Gaussian (squeezed coherent state); one can
show its Wigner distribution is a phase space Gaussian
\cite{Birk,Birkbis,Littlejohn} One can apply the same arguments as above to
discuss the effect of a variation of Planck's constant: if $\hbar^{\prime
}>\hslash$ then (\ref{RSbis}) becomes
\begin{equation}
\sigma_{x_{j}}^{2}\sigma_{p_{j}}^{2}\leq\sigma_{x_{j},p_{j}}^{2}+\tfrac{1}%
{4}\hbar^{\prime2}%
\end{equation}
and the RSI ar5e thus violated: there exists no quantum state, pure or mixed,
whose WDF is $\rho$. If we replace $\hslash$ with $\hslash^{\prime\prime
}<\hslash$ then (\ref{RSbis}) becomes%
\[
\sigma_{x_{j}}^{2}\sigma_{p_{j}}^{2}\geq\sigma_{x_{j},p_{j}}^{2}+\tfrac{1}%
{4}\hbar^{\prime\prime2}%
\]
indicating that $\rho$ is now the WDF of a mixed Gaussian state.

\section{Arbitrary Quantum States}

The dependence of arbitrary density matrices on the values of Planck's
constant is an open mathematical problem, and a\ difficult one. Consider an
arbitrary phase space function $\rho$ such that
\[
\int\rho(z)dz=1
\]
and let us ask the question: \textquotedblleft how can I know whether $\rho$
is the Wigner distribution of a (pure, or mixed) quantum
state?\textquotedblright. Defining the covariance matrix as above one can
prove (Narcowich \cite{Narcow3,Narcow}, Narcowich and O'Connell
\cite{Narconnell}) that the condition
\[
\Sigma+\frac{i\hbar}{2}J\geq0
\]
is \textit{necessary} for $\Sigma$ to be the covariance matrix of a quantum
state, but it is \textit{not sufficient}: see the discussion and the
counterexample given in de Gosson and Luef \cite{golubis}. The study of the
difficult problem to determine whether a function $\rho$ represents a quantum
state is closely related to the mathematical work of Kastler \cite{Kastler}
and Loupias and Miracle-Sole \cite{LouMiracle1,LouMiracle2} in the end of the
1960s, where the notion of function of $\hbar$-positive type was defined using
the machinery of $C^{\ast}$-algebras. This leads to a quantum version of a
classical theorem of Bochner's characterizing classical probability densities
which is difficult to use even numerically since it involves the simultaneous
verification of an uncountable infinity of inequalities. Outside Gaussian
functions, the only case which has been successfully addressed so far is that
of a pure state $|\psi\rangle$. Dias and Prata \cite{dipra} have shown, using
techniques from the theory of complex variables, that if $W\psi$ is the Wigner
distribution of a \emph{non-Gaussian} pure state $|\psi\rangle$ then
\emph{any} variation of $h$ will destroy this property. We conjecture -- but
this has yet to be proven -- that a decrease of $h$ will turn the pure state
$|\psi\rangle$ into a mixed one (as in the case of Gaussian pure states),
while an increase of $h$ will not even lead to a classical state since the
Wigner transform of a non-Gaussian function fails to be positive..

Summarizing, we are in the following situation:

\begin{itemize}
\item \textit{If we are in presence of a Gaussian state any decrease of }%
$h$\textit{ will yield a new Gaussian state; if the original state is a pure
Gaussian it will become a mixed Gaussian state. Any increase in }$h$\textit{
will transform the state into a classical (Gaussian) state};

\item \textit{A pure non-Gaussian state does not remain pure under any
variation of Planck's constant. It is not known (but however conjectured) that
it becomes a mixed state if }$h$\textit{ decreases}.
\end{itemize}

\section{Experimental Issues}

Recently one of us has shown \cite{KM1,KM3}, using GPS data, that modern
measurement techniques do not allow to prove the constancy of $h$ within an
error of 0.7\% (also see the rejoinder \cite{berengut}, and the answer
\cite{KM2}. This result opens the door to the possibility of a variable
Planck's constant. Using the sensitivity of density matrices to changes of
Planck's constant there are several possible experimental scenarios to test
this hypothesis. If one wants to study possible the dependence of $h$ on
location one of them would be the following: An atom in an excited state
propagates from B to A in a direction exactly perpendicular to the gradient of
$h$. After some time, the atom decays and emits energy. The direction of the
relaxation emission depends on the type of transition. In the case where there
is no variation in $h$, the vector momentum of the atom will change in
response to the direction and energy of the emission. In the case where there
is a variation in $h$, the atom will incur an additional deflection related to
how the emission aligns with the gradient of $h$. The size of this effect
would be vanishingly small. Meaningful detection would require high energies
and long propagation distances. High energy cosmic rays may be an avenue to
explore the validity of this prediction. This thought experiment predicts
that, if $h$ is varying, there will exist an anisotropy in the trajectories of
high energy particles undergoing a decay process that depends on their initial
velocity vector. If $h$ is not varying, but rather some dimensionless constant
such as the fine structure constant were to vary, then this anisotropy would
not exist. Another possible experimental setup is the following: imagine a
quantum state teleportation scheme that occurs between two regions that have
different values of Planck's constant. The actor \textquotedblleft
Bob\textquotedblright\ prepares a maximally entangled particle pair locally in
region B (where Planck's constant is of value $h_{B}$) and the actor
\textquotedblleft Alice\textquotedblright\ receives one of the particles and
measures it against a local Bell state prepared in region A (where Planck's
constant is of value $h_{A}$). As the transmitted particle traverses the
spacetime interval between B and A, its quantum state will modulate into a
superposition of the available local states. When the particle is received by
Alice in some final state (A), it will not be a pure Bell state in Basis set
A. However, the local half of the entangled pair held by Bob will not be
affected by this. Therefore, the teleportation process will be impossible
without some knowledge of the effective rotation operator associated with
propagation through the $\partial h/\partial z$ spacetime. This predicted
result is in direct opposition to the results expected if $h$ could not vary
between A and B, but some dimensionless constant such as the fine structure
constant could vary between A and B.

As for the case of a time-varying $h$ one could envisage the following
scenario: suppose that after the Big Bang, during the \textquotedblleft Planck
epoch\textquotedblright, where quantum theory as presently understood becomes
applicable, the Early Universe had a smaller Planck constant $h_{0}<h$ as
today. This Early Universe would then have been much more \textquotedblleft
quantum\textquotedblright\ than the current Universe; assuming\ a steady
increase of $h_{0}$ to its present value $h$ this would mean that the Universe
becomes more \textquotedblleft classical\textquotedblright\ with time. It
would be interesting to analyze (both theoretically and experimentally) what
such an evolution implies at a macroscopic level.

\section{Remarks}

The topic of \textquotedblleft variability\textquotedblright\ of physical
\textquotedblleft constants\textquotedblright, which started with Dirac's
\textit{Nature} paper \cite{Dirac} has always been a controversial one; it is
often argued that one should only test the non-constancy of dimensionless
parameters (Duff \cite{Duff}) such as, for instance, the fine-structure
constant $\alpha$ (see however the answer \cite{KM2} of one of us (MM) to the
objections in \cite{berengut}). The variability of $\alpha$ is problematic,
but many experimental results seem to point towards a non-constant value of
$\alpha$. There is however one thing which is not controversial: these are the
\emph{mathematical truths} exposed in this paper. If one accepts -- as most
quantum physicists do -- that mixed quantum states are represented by density
matrices and their Wigner distributions, there is \emph{no way} to refute the
conclusions of any experiment leading to a proof of the variability of
Planck's constant based on the mathematical dependence of states on $h$ as
exposed here.

\begin{acknowledgement}
The first author (MdG) has been financed by the Austrian Research Agency FWF;
grant number P27773--N23.
\end{acknowledgement}


\begin{thebibliography}{99}                                                                                               %


\bibitem {Barrow1}J. D. Barrow, Varying constants, Philosophical Transactions
of the Royal Society of London A: Mathematical, Physical and Engineering
Sciences 363(1834) (2005) 2139--2153.

\bibitem {BarrowWebb}J. D. Barrow and J. K. Webb, Inconstant constants: do the
inner workings of nature change with time? Sci. Am. 292(6):57--63

\bibitem {berengut}J. C. Berengut and V. V. Flambaum, Comment on
\textquotedblleft Global Positioning System Test of the Local Position
Invariance of Planck's Constant\textquotedblright, Phys. Rev. Lett.
\textbf{109}(6) (2012) 068901

\bibitem {dipra}N. C. Dias and J. N. Prata, The Narcowich-Wigner spectrum of a
pure state, \textit{Rep. Math. Phys}. \textbf{63}(1) (2009) 43--54

\bibitem {Dirac}P. A. M. Dirac, The cosmological constants, \textit{Nature}
\textbf{139} (1937) 323

\bibitem {Duff}M. J. Duff, Comment on time-variation of fundamental
constants\qquad arXiv:hep-th/0208093

\bibitem {Birk}M. de Gosson, \textit{Symplectic Geometry and Quantum
Mechanics}, Birkh\"{a}user, Basel, series \textquotedblleft Operator Theory:
Advances and Applications\textquotedblright\ Vol. \textbf{166} (2006)

\bibitem {Birkbis}M. de Gosson, \textit{Symplectic Methods in Harmonic
Analysis and in Mathematical Physics}, Birkh\"{a}user, 2011

\bibitem {golu}M. de Gosson and F. Luef, Quantum states and Hardy's
formulation of the uncertainty principle: a symplectic approach, \textit{Lett.
Math. Phys.} \textbf{80}(1) (2007) 69--82

\bibitem {golubis}M. de Gosson and F. Luef, Remarks on the fact that the
uncertainty principle does not determine the quantum state, \textit{Phys.
Lett. A} \textbf{364}(6) (2007) 453--457

\bibitem {hillery}M. O. Hillery, R. F. O'Connell, M. O. Scully, and E. P.
Wigner, Distribution functions in physics: fundamentals. InPart I: Physical
Chemistry. Part II: Solid State Physics (pp. 273--317). Springer Berlin
Heidelberg, 1997

\bibitem {Kastler}D. Kastler, The $C^{\ast}$-Algebras of a Free Boson Field,
\textit{Commun. math. Phys}. \textbf{1}, 14--48 (1965)

\bibitem {KM1}J. Kentosh and M. Mohageg, Global positioning system test of the
local position invariance of Planck's constant, Phys. Rev. Lett.
\textbf{108}(11) (2012), 110801

\bibitem {KM2}J. Kentosh and M. Mohageg, Kentosh and Mohegeg Reply,
\textit{Phys. Rev. Lett.} \textbf{109}(6) (2012) 068902

\bibitem {KM3}J. Kentosh and M. Mohageg, Testing the local position invariance
of Planck's constant in general relativity, \textit{Physics Essays
}\textbf{28}(2) (2015), 286--289

\bibitem {Littlejohn}R. G. Littlejohn, The semiclassical evolution of wave
packets, Phys. Reps. \textbf{138(}4--5) (1986) 193--291

\bibitem {LouMiracle1}G. Loupias and S. Miracle-Sole, $C^{\ast}$-Alg\`{e}bres
des syst\`{e}mes canoniques, I, \textit{Commun. math. Phys}, \textbf{2},
31--48 (1966)

\bibitem {LouMiracle2}G. Loupias and S. Miracle-Sole, $C^{\ast}$-Alg\`{e}bres
des syst\`{e}mes canoniques, II, \textit{Ann. Inst. Henri Poincar\'{e},}
\textbf{6}(1), 39--58 (1967)

\bibitem {inconstant}G. Mangano, F. Lizzi, and A. Porzio, \textit{Int. J. Mod.
Phys.} A \textbf{30}, 1550209 (2015)

\bibitem {Narcow3}F. J. Narcowich, Distributions of $\eta$-positive type and
applications, \textit{J. Math. Phys}. \textbf{30}(11), 2565--2573 (1989)

\bibitem {Narcow}F. J. Narcowich, Geometry and uncertainty, \textit{J. Math.
Phys}. \textbf{31}(2),354--364 (1990)

\bibitem {Narconnell}F. J. Narcowich and R. F. O'Connell, Necessary and
sufficient conditions for a phase-space function to be a Wigner
distribution,\textit{ Phys. Rev. A,} \textbf{34}(1), 1--6 (1986)

\bibitem {sriwolf}M. D. Srinivas and E. Wolf, Some nonclassical features of
phase-space representations of quantum mechanics, \textit{Phys. Rev. D}
\textbf{11}(6),1477--1485 (1975)

\bibitem {flambaum}J. K. Webb, M. T. Murphy, V. V. Flambaum, V. A. Dzuba, J.
D. Barrow, C. W. Churchill, J. X. Prochaska, and A. M. Wolfe, Further evidence
for cosmological evolution of the fine structure constant, \textit{Phys. Rev.
Lett}. \textbf{8}7(9) (2001) 091301

\bibitem {Werner}R. Werner, Quantum harmonic analysis on phase space,
\textit{J. Math. Phys}. \textbf{25}(5), 1404--1411 (1984).
\end{thebibliography}
\end{document}